\documentclass[%
reprint,
amsmath,amssymb,
aps,
]{revtex4-2}

\usepackage[dvipdfmx]{graphicx}
\usepackage{physics}
\usepackage{graphicx}
\usepackage{dcolumn}
\usepackage{bm}
\usepackage{color}
\usepackage{braket}
\usepackage{mathrsfs}
\usepackage{amsmath}
\usepackage{amssymb}
\usepackage{float}
\usepackage{appendix}
\usepackage{comment}

\begin{document}


\title{
Creating stars orbiting in AdS \\
}

\author{Youka Kaku${}^1$}
\author{Keiju Murata${}^2$}
\author{Jun Tsujimura${}^1$}
\affiliation{${}^1$Department of Physics, Nagoya University, Chikusa, Nagoya 464-8602, Japan}
\affiliation{${}^2$Department of Physics, College of Humanities and Sciences, Nihon University, Sakurajosui, Tokyo 156-8550, Japan}

\begin{abstract}
We propose a method to create a star orbiting in an asymptotically AdS spacetime using the AdS/CFT correspondence.
We demonstrate that by applying an appropriate source in the quantum field theory defined on a 2-sphere, the localized star gradually appears in the dual asymptotically AdS geometry.
Once the star is created, the angular position can be observed from the response function.
The relationship between the parameters of the created star and those of the source is studied. 
We show that information regarding the bulk geometry can be extracted from the observation of stellar motion in the bulk geometry.
\end{abstract}


\maketitle


\setcounter{footnote}{0}

\noindent
{\it Introduction.}--- 
Stellar motion around Sagittarius A$^\ast$ has been observed for decades, 
and these observations provide strong evidence for the existence of a black hole at the centre of our galaxy~\cite{GRAVITY:2018ofz}.
They have also provided important information regarding the curved spacetime around the black hole.
In this letter, we propose a method for creating a star orbiting in an asymptotically AdS spacetime using the AdS/CFT correspondence~\cite{Maldacena:1997re,Gubser:1998bc,Witten:1998qj}.
We also discuss how it is possible to extract information about the bulk geometry from the stellar motion.
Our main target is AdS/CFT in the ``bottom-up approach'', 
such as the correspondence between condensed matter systems and gravitational systems~\cite{Hartnoll:2009sz,Herzog:2009xv,McGreevy:2009xe,Horowitz:2010gk,Sachdev:2010ch}. 
In many cases, there is no concrete guiding principle for constructing dual gravitational theories of condensed matter.
Our proposal provides a direct way to extract information regarding the dual geometries of condensed matter through experiments.

\begin{figure}
\begin{center}
\includegraphics[scale=0.3]{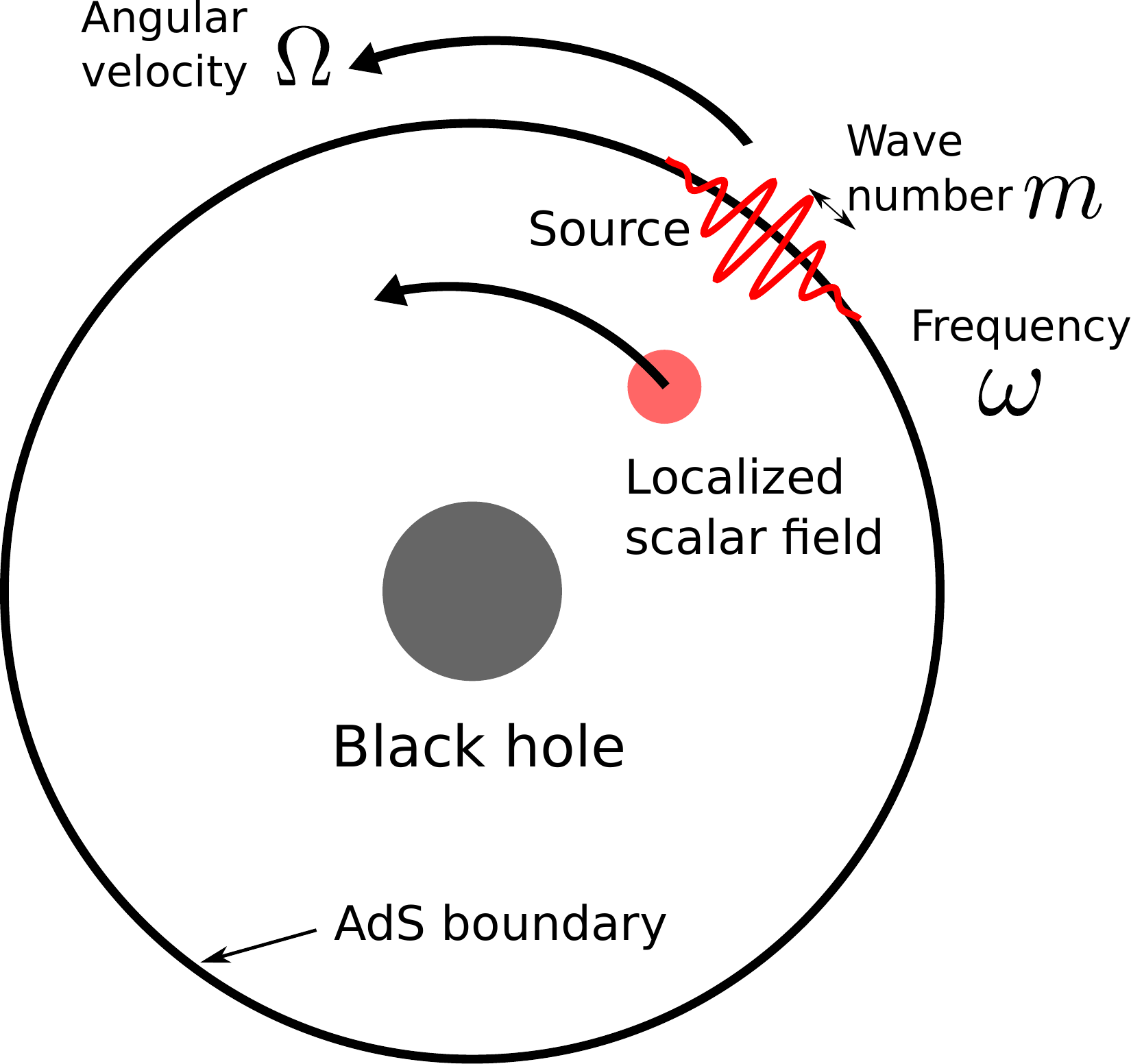}
\end{center}
\caption{
Schematic image of our setup.
}
\label{ponchie}
\end{figure}

Fig.\ref{ponchie} shows a schematic image of our setup. 
We consider the pure global AdS and Schwarzschild-AdS$_4$ (Sch-AdS$_4$) spacetime with a spherical horizon as the background spacetimes.
These correspond to the ($2+1$)-dimensional quantum field theory (QFT) on $S^2$. We  deal with the bulk scalar field as the probe, which corresponds to a scalar operator $\mathcal{O}$ in the dual QFT.
We regard the operator $\mathcal{O}$ as the source of the bulk field.  The source is localized in $S^2$, and its packet rotates with angular velocity $\Omega$.
It also has frequency $\omega$ and wavenumber $m$. 
We demonstrate that by tuning the parameters $(\omega,m,\Omega)$, a bulk star is created.

Previous studies have proposed that gravitational lensing can be used to test the existence of a given QFT~\cite{Hashimoto:2018okj,Hashimoto:2019jmw,Kaku:2021xqp,Zeng:2022woh}.
The Einstein ring formed by gravitational lensing provides information about the photon sphere of the null geodesic in dual geometry. 
In this letter, we propose another method to probe dual geometry using the timelike geodesic. 
In Refs.\cite{Berenstein:2019tcs,Berenstein:2020vlp}, dual operators corresponding to localised states in the AdS bulk have been investigated. 
Our work provides an explicit source function for creating similar states through a time evolution.

$\,$\\
{\it Eikonal approximation for massive scalar field}--- 
We consider the Sch-AdS$_4$ with the spherical horizon as
\begin{align}
ds^2 = -f(r)dt^2 + \frac{dr^2}{f(r)} + r^2\qty( d\theta^2 + \sin^2\!\theta\, d\phi^2 ) ,
\end{align}
where $f(r)=1 + r^2 - r_h(1+r_h^2)/r$ in units of the AdS radius. 
For $r_h=0$, this spacetime describes the pure global AdS.
Let us consider the circular orbit of the massive particle in this spacetime. 
The specific energy and angular momentum of the particle are given by
$\epsilon \equiv -u_t$ and $j\equiv u_\phi$, where $u^\mu$ is the 4-velocity. 
The angular velocity of revolution is $\Omega\equiv d\phi/dt=u^\phi/u^t$.
For a circular orbit with radius $r=R$, the parameters of the timelike geodesic ($\epsilon$, $j$, $\Omega$) are given by the one-parameter family of $R$ (for fixed $r_h$) as
\begin{equation}
\begin{split}
&\epsilon^2 = \frac{2 (R-r_h)^2 (R^2+r_h R + r_h^2 + 1)^2}{R\{2R-3 r_h (1+r_h^2))\}}\ , \\
&j^2 = \frac{R^2 (2 R^3 + r_h^3 + r_h)}{2R-3 r_h (1+r_h^2)}\ ,\ 
\Omega^2 = \frac{2 R^3 + r_h^3 + r_h}{2R^3}\ .
\end{split}
\label{paras_circ}
\end{equation}
We will consider the creation of the massive particle (or star) 
as the coherent excitation of the bulk field. 

We deal with the massive scalar field in a fixed background whose Lagrangian is given by:
\begin{equation}
\mathcal{L}=-(\partial\Phi)^2-\mu^2 \Phi^2\ .
\label{scalarL}
\end{equation}
The scalar field obeys the Klein-Gordon equation $\square \Phi=\mu^2 \Phi$.
Using the Eikonal approximation, we can obtain the timelike geodesic equation from the Klein-Gordon equation.
We assume that the typical frequency $\omega$ and mass $\mu$ of the scalar field are sufficiently large, and that they are of the same order, $\omega\sim\mu\gg 1$.
Substituting 
$\Phi(x^\mu)=a(x^\mu) e^{iS(x^\mu)}$
into the Klein-Gordon equation and assuming $\partial_\mu S \sim \mathcal{O}(\omega)$,
we obtain 
\begin{equation}
g^{\mu\nu}\partial_\mu S \partial_\nu S = -\mu^2
\end{equation}
as the leading-order equation for $\omega$. 
Introducing the 4-velocity $u_\mu = \partial_\mu S/\mu$, we have $u_\mu u^\mu = -1$. 
Differentiating this equation, we also obtain the geodesic equation for a massive particle as 
$0=\nabla_\rho(g^{\mu\nu}\partial_\mu S \partial_\nu S)/\mu^2=2u^\mu \nabla _\mu u_\rho$.
Thus, the relationship between the parameters of the timelike geodesic and the massive scalar field is
\begin{equation}
\epsilon = -\frac{1}{\mu}\partial_t S \ ,\quad
j = \frac{1}{\mu}\partial_\phi S \ .
\label{ejrel0}
\end{equation}
Analysis of the Eikonal approximation indicates that massive particles should also be expressed as the localized configuration of the massive scalar field.
Our main task is to determine the appropriate boundary condition for the scalar field at the AdS boundary and create a particle (or star) 
orbiting in AdS, as shown in Fig.\ref{ponchie}.

$\,$\\
{\it Massive scalar field in asymptotically AdS spacetimes}--- 
Near the AdS boundary $r=\infty$, the scalar field behaves as
\begin{equation}
\Phi(t,r,\theta,\phi)\simeq \mathcal{J}(t,\theta,\phi)\,r^{-\Delta_-} + \langle \mathcal{O}(t,\theta,\phi)\rangle\, r^{-\Delta_+}\ ,
\label{JOdef}
\end{equation}
where $\Delta_\pm = 3/2\pm \nu$ and $\nu=\sqrt{9/4+\mu^2}$. 
We refer to $\mathcal{J}(t,\theta,\phi)$ and $\langle \mathcal{O}(t,\theta,\phi)\rangle$ as the ``source'' and ``response'', respectively.

Caution is needed when considering the ``source'' for the massive scalar field.
For $\mu^2>0$, the corresponding operator $\mathcal{O}$ has a conformal weight $\Delta_+>3$, and
applying the source to such an operator corresponds to an irrelevant deformation of the dual QFT.
From a gravitational point of view, if a non-asymptotically mode is present, 
the energy-momentum tensor of the scalar field diverges near the AdS boundary and the probe approximation is no longer valid ~\cite{vanRees:2011fr}.
One way to avoid this problem is to introduce an explicit cutoff at the finite radius of asymptotically AdS spacetime.
The AdS with a finite radial cutoff is considered the gravitational dual of the $T\bar{T}$-deformed theory~\cite{McGough:2016lol}.
The other way is to introduce a renormalization group flow to a UV fixed point where $\mathcal{O}$ is relevant. 
One of the simplest examples is the addition of another scalar field $\psi$, which controls the mass for $\Phi$: 
\begin{equation}
\mathcal{L}'=-(\partial\Phi)^2-\lambda(\psi)^2 \Phi^2 -(\partial\psi)^2+2\psi^2\ ,
\label{scalarLp}
\end{equation}
where $\lambda(\psi)$ is now a function of the dynamic scalar field $\psi$. 
Because the mass square of $\psi$ is $-2$, $\psi$ behaves as $\psi\sim 1/r, 1/r^2$ near the AdS boundary, and both modes are normalizable.
When $\Phi=0$, we can obtain a static and spherically symmetric profile $\psi=\psi(r)$ by imposing only the regularity at the horizon or centre of the global AdS. 
(Then, both the $1/r$ and $1/r^2$ modes are present at the AdS boundary in general.)
By carefully choosing the mass function $\lambda(\psi)$, for example, $\lambda(\psi)=\mu \tanh(\psi)$, the effective mass for $\Phi$ can be almost constant, except for the region near the AdS boundary.
Considering the infinitesimal perturbation of $\Phi$, we have a scalar field whose mass vanishes near the AdS boundary and whose energy-momentum tensor is still finite.
(The backreaction to $\psi$ is second-order in $\Phi$ and negligible.)
Let us take the cutoff $r=\Lambda$ so that
\begin{equation}
\lambda(\psi) = 
\begin{cases}
\mu & (r\lesssim \Lambda)\\
0 & (r\gtrsim \Lambda)
\end{cases}
\ ,
\end{equation}
is satisfied.
We consider only the region $r\lesssim \Lambda$, where the theory is described by Eq.(\ref{scalarL}).
Subsequently, the ``source'' in Eq.(\ref{JOdef}) can be regarded as $\mathcal{J}\simeq r^{\Delta_-}\Phi|_{r=\Lambda}$.
Although this $\mathcal{J}$ is different from the ``real'' source defined in UV-complete theory~(\ref{scalarLp}), 
$\mathcal{J}_\textrm{UV}=\Phi|_{r=\infty}$, 
we assume that they are qualitatively similar because $\Phi(r=\Lambda)$ and $\Phi(r=\infty)$ are only related to the $r$ evolution of the equation of motion derived from Eq.(\ref{scalarLp}).
In this letter, we refer to $\mathcal{J}$ as the source.

$\,$\\
{\it Massive scalar field localized in bounded orbit.}--- 
We adopt the following form of the source function:
\begin{multline}
\mathcal{J}(t,\theta,\phi) = J_0 \, \exp\bigg[-\frac{(t-T)^2}{2\sigma_t^2}\\
-\frac{(\theta-\pi/2)^2}{2\sigma_\theta^2}
-\frac{(\phi-\Omega t)^2}{2\sigma_\phi^2}
-i\omega t+im\phi \bigg]\ .
\label{src}
\end{multline}
This function is localized in $S^2$ at $\theta=\pi/2$ and $\phi=\Omega t$, with widths $\sigma_\theta$ and $\sigma_\phi$, respectively. 
(We take the domain of the coordinate $\phi$ as $-\pi<\phi-\Omega t \leq \pi$.)
The centre of the localized source rotates on the equator with angular velocity $\Omega$. 
This has a wavenumber $m$ along the $\phi$-direction and oscillates over time with frequency $\omega$. 
(See Fig.\ref{ponchie} for the schematic picture of the source.)
The source is also localized in time at $t=T$ with width $\sigma_t$. 
We take a large $\sigma_t$ such that the modes with frequency $\sim \omega$ are sufficiently excited.
We take $T<0$ and $\sigma_t\ll |T|$ such that the application of the source has already been terminated at $t=0$.
$J_0$ is the amplitude of the source; however, it is not important in our analysis because of the linearity of the scalar field.

There are some requirements for the parameters in Eq.(\ref{src}) to realise a localized star in the bulk.
The scalar field induced by the source~(\ref{src}) typically has a frequency $\omega$ and wavenumber $m$.
In addition, its angular size is determined by $\sigma_\theta$ and $\sigma_\phi$. 
Conversely, in momentum space, the scalar field is distributed with a width $\sim 1/\sigma_\phi,1/\sigma_\theta$. 
Therefore, the condition that the scalar field is localized in both the real and momentum spaces is given by:
\begin{equation}
\frac{1}{m} \ll \sigma_\theta, \sigma_\phi \ll 1\ .
\end{equation}
It is also necessary that $\omega$ and $\mu$ be sufficiently large compared with the curvature scale of the bulk geometry, 
so that the Eikonal approximation is valid.

Eq.(\ref{src}) is regarded as the boundary condition of the scalar field near the AdS boundary. 
We now explain how this boundary condition is imposed and the equation of motion for the scalar field is solved.
If we decompose $\Phi$ as $\Phi =r^{-1} \sum_{\omega' m' l'} e^{-i\omega' t} \Psi_{\omega' l' m'}(x)Y_{l' m'}(\theta,\phi)$, where $Y_{l'm'}$ is the spherical harmonics, then $\Psi_{\omega l'm'}(x)$ obeys the equation in 
the Schr\"{o}dinger form:
\begin{gather}
\label{eq:Klein-Gordon}
\qty[ -\frac{d^2}{dx^2}+ V(x)] \Psi_{\omega' l'm'}(x) = \omega'{}^2 \Psi_{\omega' l'm'}(x), \\
\label{eq:Potential}
V(x) = f(r)\qty( \frac{l'(l'+1)}{r^2}+ \mu^2 +\frac{1}{r}\frac{d f}{dr} ),
\end{gather}
where $x= \int dr/f(r)$ is the tortoise coordinate. 
We can also decompose the source~(\ref{src}) as
\begin{equation}
\mathcal{J}(t,\theta,\phi)=\sum_{\omega' l' m'} J_{\omega' l' m'}\, e^{-i\omega' t} Y_{l' m'}(\theta,\phi)\ .
\end{equation}
The coefficient $J_{\omega' l' m'}$ provides the boundary condition for $\Psi_{\omega' l' m'}(x)$ at the AdS boundary: 
$\Psi_{\omega' l' m'}(x)\to J_{\omega' l' m'} r^{-\Delta_- +1}$.
For $r_h>0$, we impose the ingoing wave boundary condition at horizon $\Psi_{\omega' l' m'}(x) \sim e^{-i\omega' x}$. 
For $r_h=0$, we impose regularity at the centre of the AdS, $\Psi_{\omega' l' m'}(x)\sim r^{l'}$.
Under these boundary conditions, Eq.(\ref{eq:Klein-Gordon}),
and superposing the numerically obtained solutions, 
we obtain the scalar field in real space $\Phi(t,r,\theta,\phi)$. (See the supplementary material for details). 

For source~(\ref{src}), the typical frequency and wavenumber of the bulk scalar field are given by $\omega$ and $m$, respectively. 
From Eq.(\ref{ejrel0}), the specific energy and angular momentum of the created star are given by
\begin{equation}
\epsilon =\frac{\omega}{\mu} \ ,\quad
j = \frac{m}{\mu} \ .
\end{equation}
We can expect the angular velocity of the revolution of the star to be determined by $\Omega$ in Eq.(\ref{src}). 
This is verified by our numerical results. 
The rest mass is the energy measured by the observer accompanying the star: $\sim \int d\Sigma T_{\mu\nu} u^\mu u^\nu$, 
where $T_{\mu\nu}$ is the energy-momentum tensor of the scalar field and $\int d\Sigma$ denotes the integral on the $t=$const surface. 
This is proportional to $|J_0|^2$ with other parameters fixed.
The relationships between the parameters of the created star and those of the external source are summarized in Table \ref{tab:table_src_geod}.
The scalar field is localized at the local minimum of the effective potential $V(x)$. 
The radial size is determined by its curvature $\sigma_x=(V_{,xx}|_\textrm{local min})^{-1/2}$.

\begin{table}[h]
\caption{\label{tab:table_src_geod}%
Relationship between parameters of the created star and those of the external source $\mathcal{J}(t,\theta,\phi)$.
}
\begin{ruledtabular}
\begin{tabular}{ l c }
\textrm{Physical quantities of star}&
\textrm{Parameters of source}\\
\colrule
Specific energy & $\omega/\mu$\\
Specific angular momentum & $m/\mu$\\
Rest mass & $\propto |J_0|^2$\\
Angular velocity of revolution & $\Omega$\\
Size & $\sigma_\theta,\sigma_\phi,\sigma_x=(V_{,xx}|_\textrm{local min})^{-1/2}$
\end{tabular}
\end{ruledtabular}
\end{table}

Note that Table \ref{tab:table_src_geod} does not mean that a star is created for any value of $\omega$, $m$, and $\Omega$.
As in Eq.(\ref{paras_circ}), $\epsilon$, $j$, and $\Omega$ are given by the one-parameter family of the radius of circular orbit $R$.
Hence, if we want to create a star at $r=R$, we need to tune $\omega$, $m$, and $\Omega$ to the values obtained by these equations and Table \ref{tab:table_src_geod}.

$\,$\\
{\it Results.}--- 
\begin{figure*}[t]
\centering
\includegraphics[width=150mm]{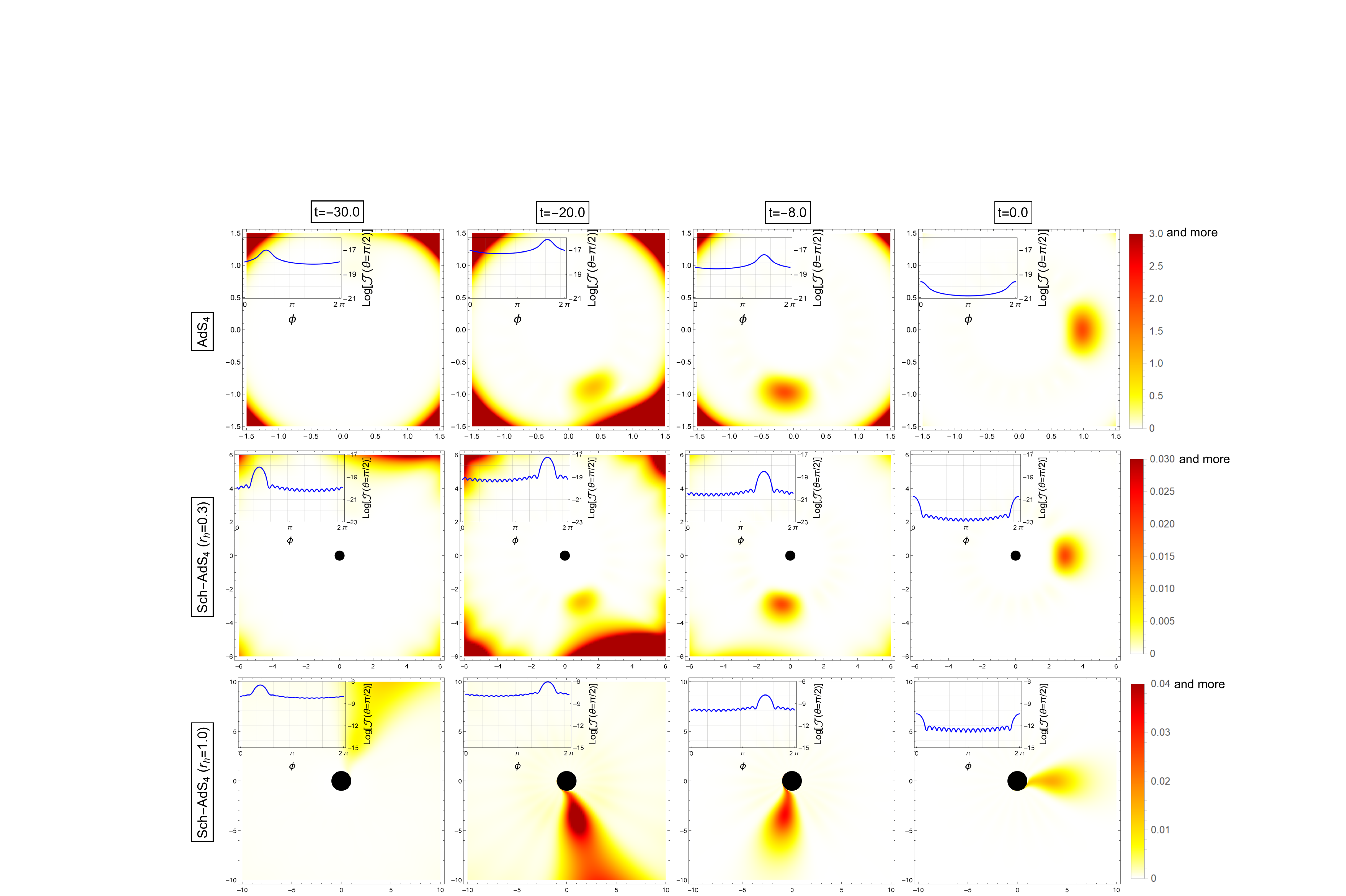}
\caption{Time evolution of the creation of the scalar field orbiting in AdS$_4$ and Sch-AdS$_4$. 
(Animated gifs are available in ancillary files of arXiv.)
}
\label{fig:timelapse}
\end{figure*}
In Fig.~\ref{fig:timelapse}, we depict the time evolution of the scalar field orbiting in the equatorial plane $\theta=\pi/2$. 
The parameters used in our numerical calculation are summarized in Table \ref{tab:src_parameters}. 
\begin{table}[hbtp]
\caption{Parameters for numerical calculations, where $\nu^2\equiv \mu^2+9/4$.}
\centering
\begin{tabular}{|c|cccc|cc|}
\hline
$r_h$ & $\nu$ &$\omega$ & $m$ & $\Omega$ & $\epsilon=\omega/\mu$ & $j=m/\mu$\\
\hline
0 & 50 & 101.5 & 50 & 1 & 2.0309 & 1.0005 \\
0.3 & 20.5 & 230.78 & 210 & 1.00312 & 11.288 & 10.271 \\
1 & 5.5 & 215.19 & 210 & 1.00727 & 40.667 & 10.271 \\
\hline
\end{tabular}
\label{tab:src_parameters}
\end{table}
In addition, we set $\sigma_\theta=\sigma_\phi=0.2, T=-20, \sigma_t=5$. 
The black disks are the event horizons, and each embedded figure on the top left shows the source~(\ref{src}) at $\theta=\pi/2$ with respect to $\phi$.
The scalar field accumulates at the local minimum of the potential~(\ref{eq:Potential}) and forms a localized wave packet revolving anticlockwise.
Using Table \ref{tab:table_src_geod}, we can estimate the specific energy $\omega/\mu$ and angular momentum $m/\mu$ of the corresponding timelike geodesic.
From Eq.(\ref{ejrel0}), the radii of revolution are $R\simeq 1.015, 3.07, 5.02$ for $r_h=0,0.3,1$, respectively.
This was consistent with the results shown in Fig.~\ref{fig:timelapse} and indicates that the motion of the created star obeys the timelike geodesic equation.

\begin{figure}[H]
\centering
\includegraphics[width=60mm]{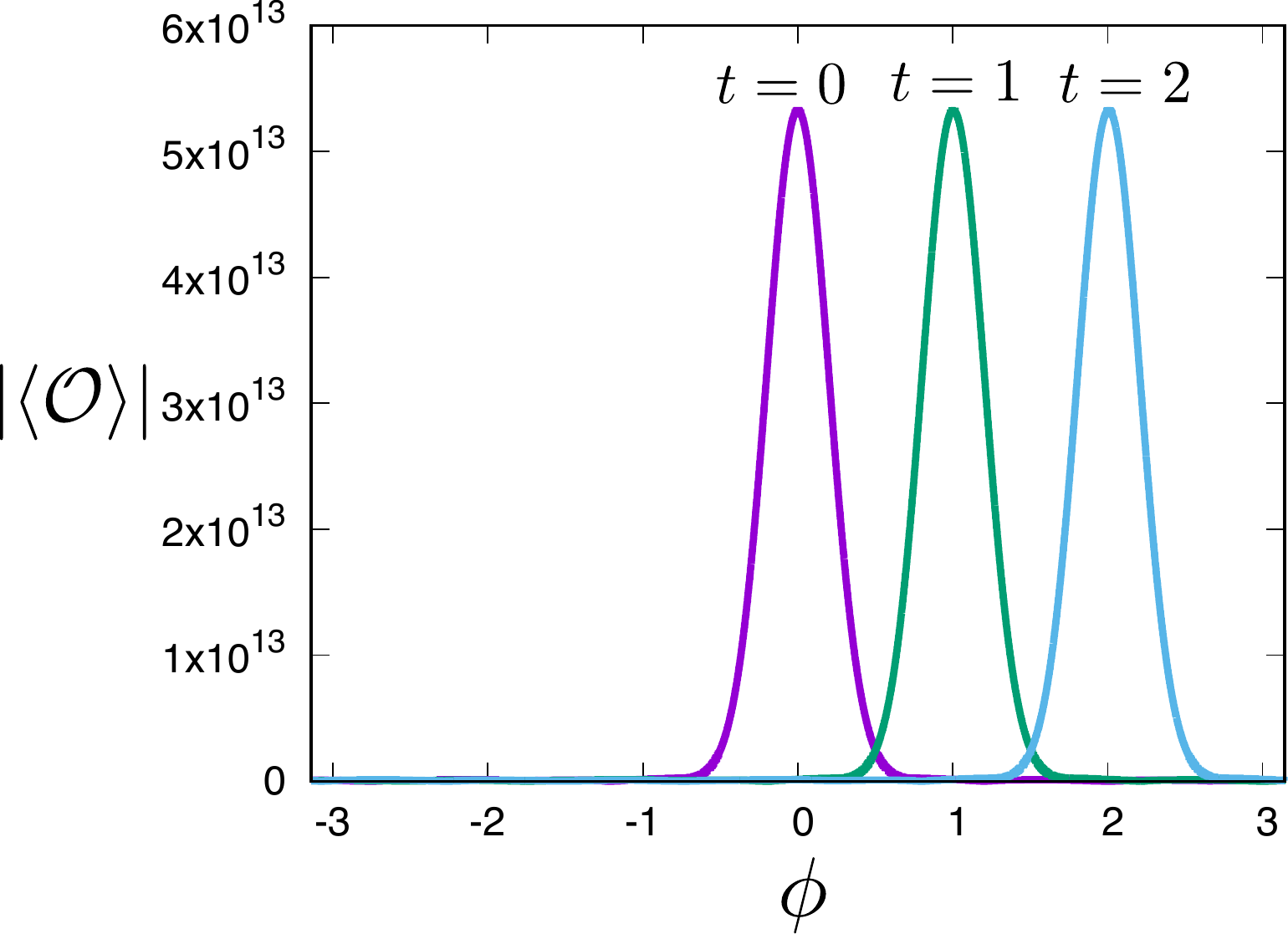}
\caption{Response in Sch-AdS$_4$ ($r_h=0.3, \nu=20.5, M=210, \sigma_\theta=\sigma_\phi=0.2, T=-20, \sigma=5$)}
\label{fig:response}
\end{figure}

Generally, in Sch-AdS$_4$ spacetime, the amplitude of the scalar field decays in time because of tunnelling towards the horizon.
The decay rate is characterised by the imaginary part of the quasi-normal mode frequency $\omega_\textrm{qnm}$. 
For $r_h=0.3$, the potential barrier is high and the decay rate is extremely suppressed.
Conversely, for $r_h=1.0$, we have $\omega_\textrm{qnm}\simeq 215 - 0.0932 i$ for $l'=210$, and the time scale of the decay is $\tau_\text{decay}=10.7$.
This is why the scalar field decays at a later time in the bottom line of Fig.~\ref{fig:timelapse}]. 
Although we employed modest values for $\omega,m$ and $\mu$ because of the limitations of computational power, 
in principle, we can realise a long-lived localized scalar field by using larger values for $\omega,m$ and $\mu$ for fixed $\omega/\mu$ and $m/\mu$.
Thus, a higher potential barrier is realized, and we have a small decay rate. 
Once we obtain the star orbiting in an asymptotically AdS spacetime, we can compute the response function from Eq.(\ref{JOdef}).
In Fig.~\ref{fig:response}, we depict the response of $\theta=\pi/2$ for $r_h=0.3$ after the star is created $t\gtrsim 0$. 
The response circulates on the equator, following the orbiting scalar field. This indicates that the angular position of the star can be observed using the response function.

$\,$\\
{\it Discussion.}--
We demonstrated that a star orbiting in the asymptotically AdS spacetime can be created 
by applying the appropriate source~(\ref{src}) in the dual QFT.
The parameters in the source should be tuned to create the localized star. 
If the dual geometry is known, 
we can determine the parameters by studying the timelike geodesic, as in Eq.(\ref{ejrel0}).
However, for a real material, in general, we do not know the dual geometry explicitly.
Thus, in a real experiment, 
we must tune the parameters $\omega$, $m$, and $\Omega$ by trial and error. 
The creation of a star in the bulk is verified by the response function, as shown in Fig.\ref{fig:response}.
Once we can create a star in the bulk, we obtain the relationship between $\epsilon$, $m$, and $\Omega$: $j=j(\epsilon)$ and $\Omega=\Omega(\epsilon)$. 
This provides information regarding the geometry of the AdS bulk.

In this letter, only a circular orbit was considered: 
the scalar field was radially localised at the local minimum of the potential~(\ref{eq:Potential}).
The non-circular orbit is the coherent excitation around the local minimum.
Such a bulk coherent state can be realized by varying $J_0$ in time in the source~(\ref{src}).
Observing the star in the non-circular orbit through the response function, 
we can obtain information about a wider region of the bulk geometry.

\begin{acknowledgments}
We would like to thank 
Takaaki Ishii and Kotaro Tamaoka
for useful discussions and comments.
This work of K.M. was supported in part by JSPS KAKENHI Grant Nos. 20K03976, 18H01214, and 21H05186. 
This work of J.T. was financially supported by JST SPRING, Grant Number JPMJSP2125. 
The author J.T. would also like to take this opportunity to thank the ``Interdisciplinary Frontier Next-Generation Researcher Program of the Tokai Higher Education and Research System.''
We would like to thank Editage (www.editage.com) for English language editing.
\end{acknowledgments}


\bibliography{MassiveStar}

\end{document}



\title{
Supplemental material for Creating stars orbiting in AdS \\
}

\author{Youka Kaku${}^1$}
\author{Keiju Murata${}^2$}
\author{Jun Tsujimura${}^1$}
\affiliation{${}^1$Department of Physics, Nagoya University, Chikusa, Nagoya 464-8602, Japan}
\affiliation{${}^2$Department of Physics, College of Humanities and Sciences, Nihon University, Sakurajosui, Tokyo 156-8550, Japan}

\begin{abstract}
In this supplemental material, 
we provide a practical way to find the boundary condition for the scalar field to create a star orbiting in asymptotically AdS spacetime.
We also describe the numerical method used to determine the time evolution of the scalar field for a given boundary condition.
\end{abstract}


\maketitle


\setcounter{footnote}{0}

\section{Klein-Gordon equation in Schwarzschild-AdS$_4$}
\label{KGAdS}

We consider the Klein-Gordon equation $\square \Phi=\mu^2 \Phi$ in Schwarzschild-AdS$_4$ with a spherical horizon:
\begin{align}
ds^2 = -f(r)dt^2 + \frac{dr^2}{f(r)} + r^2\qty( d\theta^2 + \sin^2\!\theta\, d\phi^2 ) ,
\end{align}
where $f(r)=1 + r^2 - r_h(1+r_h^2)/r$ is the AdS radius and $r_h$ is the horizon radius.
Our main goal is to create a localized profile of the scalar field, which is orbiting in the AdS spacetime.

\begin{figure}[htbp]
\includegraphics[width=80mm]{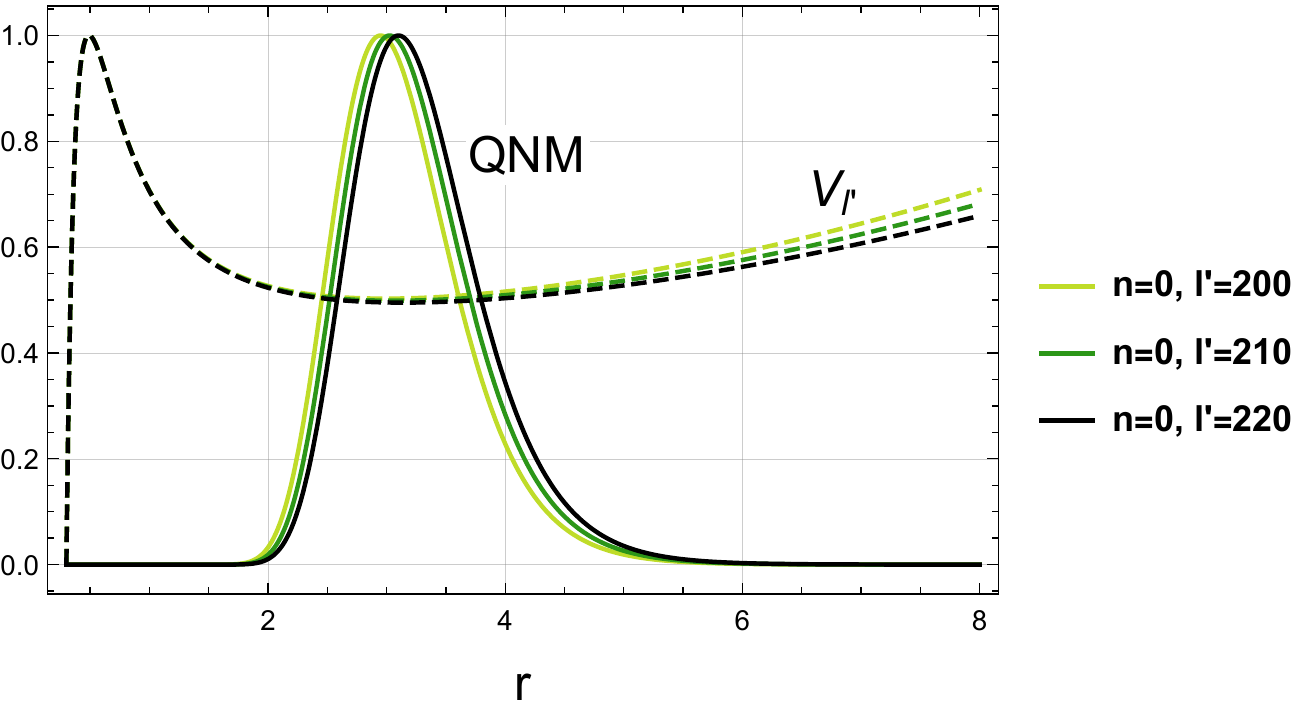}
\caption{Effective potentials of the scalar field in Sch-AdS$_4$. 
The parameters are $r_h=0.3, \nu=20.5, n=0, l'=200,210,220$. The fundamental quasi-normal mode is localized at the local minimum potential.}
\label{fig:VandQNM}
\end{figure}

Decomposing the scalar field by spherical harmonics as $\Phi(t,r,\theta,\phi)=r^{-1}\Psi_{l'm'}(t,r)Y_{l'm'}(\theta,\phi)$,
we obtain the wave equation in two dimensions as
\begin{align}
\label{eq:schrodinger}
\left(-\partial_t^2+\partial_x^2-V_{l'}(x)\right)\Psi_{l'm'}(t,r)=0,
\end{align}
where 
\begin{equation}
x\equiv\int^r_{r_0} \frac{dr'}{f(r')}\ ,
\end{equation}
is a tortoise coordinate system. We consider $r_0=0$ and $r_0=\infty$ for $r_h=0$ and $r_h>0$, respectively.
The effective potential is given by
\begin{align}
V_{l'}(x) = f(r)\qty( \frac{l'(l'+1)}{r^2}+ \mu^2 +\frac{1}{r}\frac{d f}{dr} )\ .
\label{eq: Schrodingereq}
\end{align}
Near the AdS boundary $r=\infty$, the scalar field behaves as
\begin{equation}
r^{-1}\Psi_{l'm'}(t,r)\simeq J_{l'm'}(t)\,r^{-\Delta_-} 
+ \langle \mathcal{O}_{l'm'}(t)\rangle\, r^{-\Delta_+}\ ,
\end{equation}
where $\Delta_\pm = 3/2\pm \nu$ and $\nu=\sqrt{9/4+\mu^2}$.
The typical profile of the effective potential is shown in Fig.\ref{fig:VandQNM}.
For appropriate parameters, the potential has a local minimum. 
The radially localized solution at the local minimum is given by the fundamental quasi-normal mode, which is shown in Fig.\ref{fig:VandQNM}.
Our first task is to dynamically create this localized quasi-normal mode by choosing the ``source'' $J_{l'm'}(t)$.

\section{Toy example: String vibration}

\begin{figure}[htbp]
\includegraphics[width=80mm]{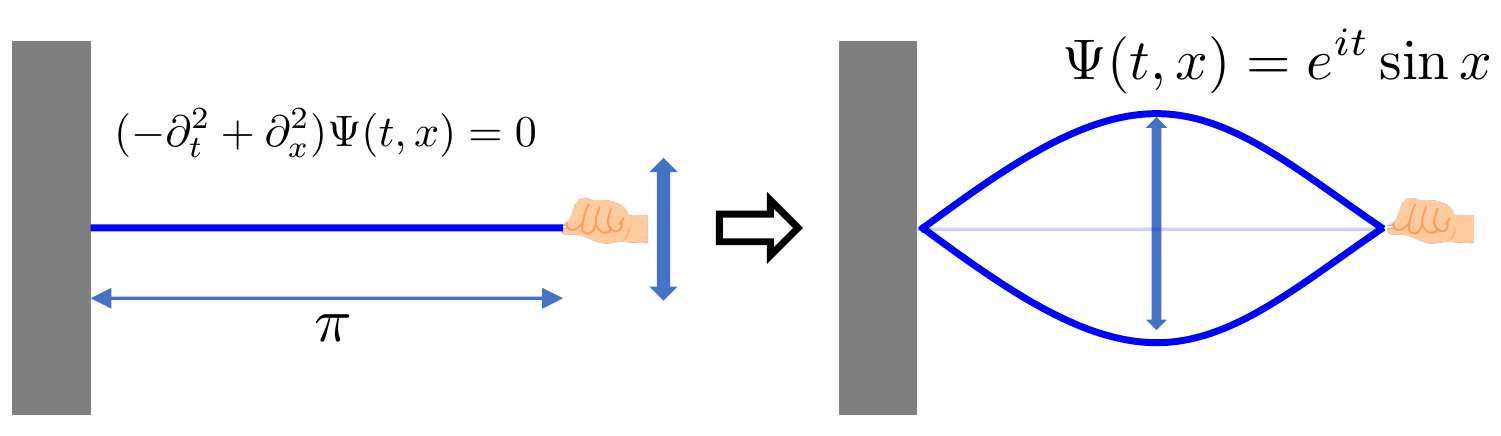}
\caption{
Illustration of our toy example. }
\label{fig:shake}
\end{figure}

Let us consider string vibration as a toy example.
Our setup is illustrated in Fig.\ref{fig:shake}: 
the string is initially static, and we can shake the endpoint of the string as desired. 
How should we shake the endpoint to realize a single normal mode at a later time?
This is a good exercise to understand the problem raised in the previous section. 

The amplitude of the string oscillation obeys an ordinary wave equation:
\begin{equation}
(-\partial_t^2+\partial_x^2) \Psi(t,x)=0\ .
\label{waveeq}
\end{equation}
We take its domain as $0\leq x \leq \pi$ and impose the Dirichlet boundary conditions
\begin{equation}
\Psi(t,x=0)=0\ ,\quad \Psi(t,x=\pi)=J(t)\ .
\label{wavebc}
\end{equation}
We assume that the endpoint is not shaken sufficiently early or late: $J(t)\to 0$, for $t\to \pm \infty$.
As the initial condition, we impose
\begin{equation}
\Psi(t,x)|_{t\to-\infty}=0\ .
\label{waveini}
\end{equation}
In the case of $J(t)=0$, we have normal modes
\begin{equation}
\Psi_n(t,x)=e^{-i n t} \sin nx\ ,
\end{equation}
whose eigenfrequencies are $\omega_n=n$ for $n=1,2,3,\cdots$.
Let us seek function $J(t)$ to realize $\Psi_n(t,x)$ at a sufficiently late time.

\begin{figure}[htbp]
\includegraphics[width=60mm]{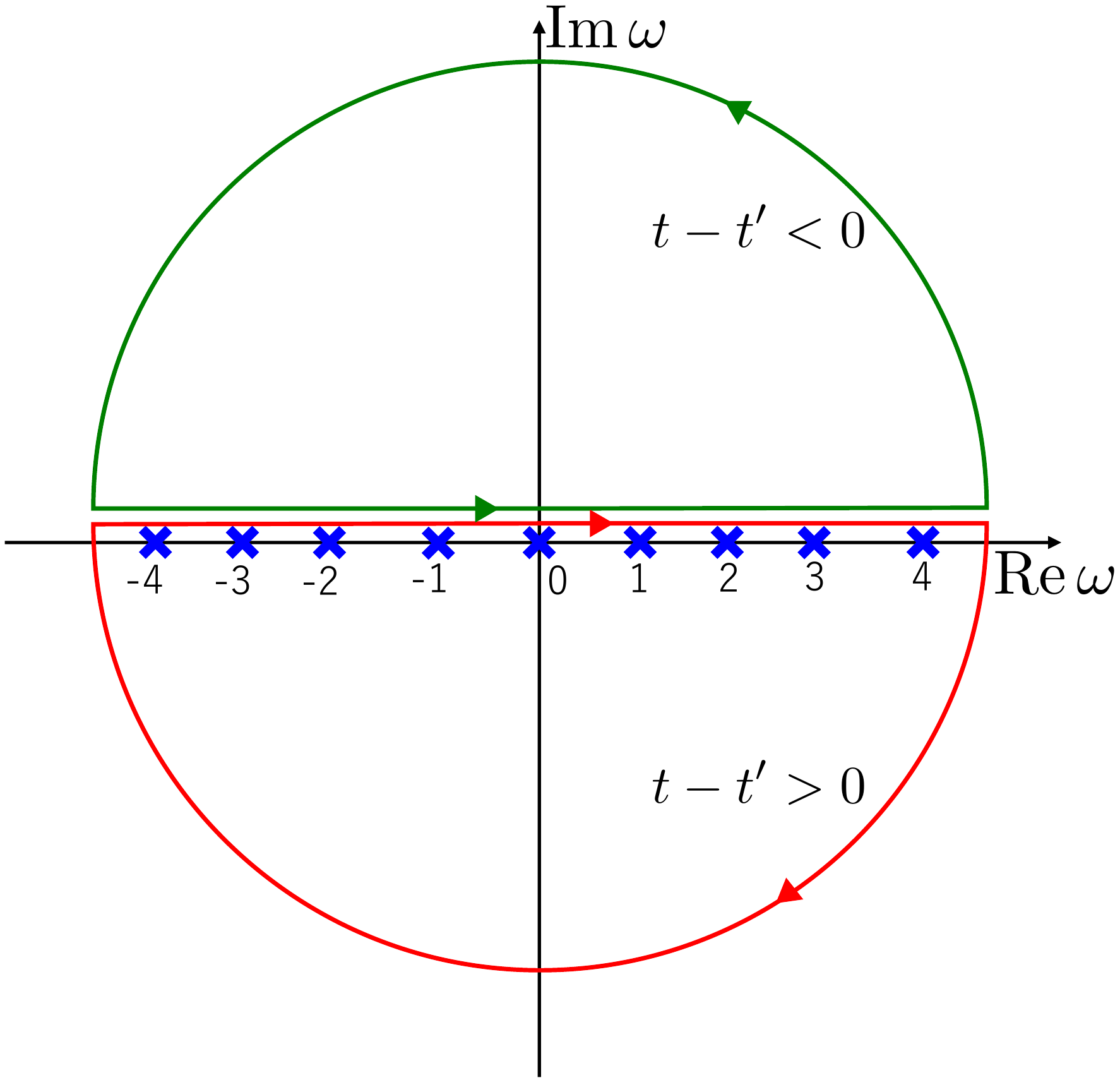}
\caption{Contour of the $\omega$ integration in Eq.(\ref{Psitoy})}
\label{fig:contour0}
\end{figure}

We can express the general solution of Eq.\eqref{waveeq} as 
\begin{align} 
\Psi(t,x) = \int_{-\infty}^{\infty} \frac{d\omega}{2\pi} C(\omega) e^{-i \omega t} \sin \omega x \ ,
\label{eq:sin_integral0}
\end{align}
This satisfies $ \Psi (t, x = 0) = 0 $ trivially. Applying the Fourier transformation to the boundary condition of the other side, $\Psi(t,x=\pi)=J(t)$, we obtain:
\begin{equation}
C(\omega) = \frac{\tilde{J}(\omega)}{\sin \pi \omega}\ ,
\end{equation}
where $\tilde{J}(\omega)$ is the Fourier transformation of $J(t)$, i.e,
\begin{equation}
\tilde{J}(\omega) \equiv \int_{-\infty}^{\infty} dt'\, J(t') e^{i \omega t'}\ .
\label{Jtilde}
\end{equation}
Thus, we find the solution of the wave equation satisfying Eq.(\ref{wavebc}) as
\begin{align} 
\Psi(t,x) = \int_{-\infty}^{\infty} \frac{d\omega}{2\pi} \frac{\tilde{J}(\omega)}{\sin \pi \omega} e^{-i \omega t} \sin \omega x \ .
\label{eq:sin_integral}
\end{align}
Note that the integrand has an infinite number of poles at normal mode frequencies $\omega=n$.
We consider the integral contour that passes through the upper side of the poles (See Fig.\ref{fig:contour0}.) 
This contour corresponds to the initial condition~(\ref{waveini}), as we will see shortly.

\begin{figure}[htbp]
\includegraphics[width=65mm]{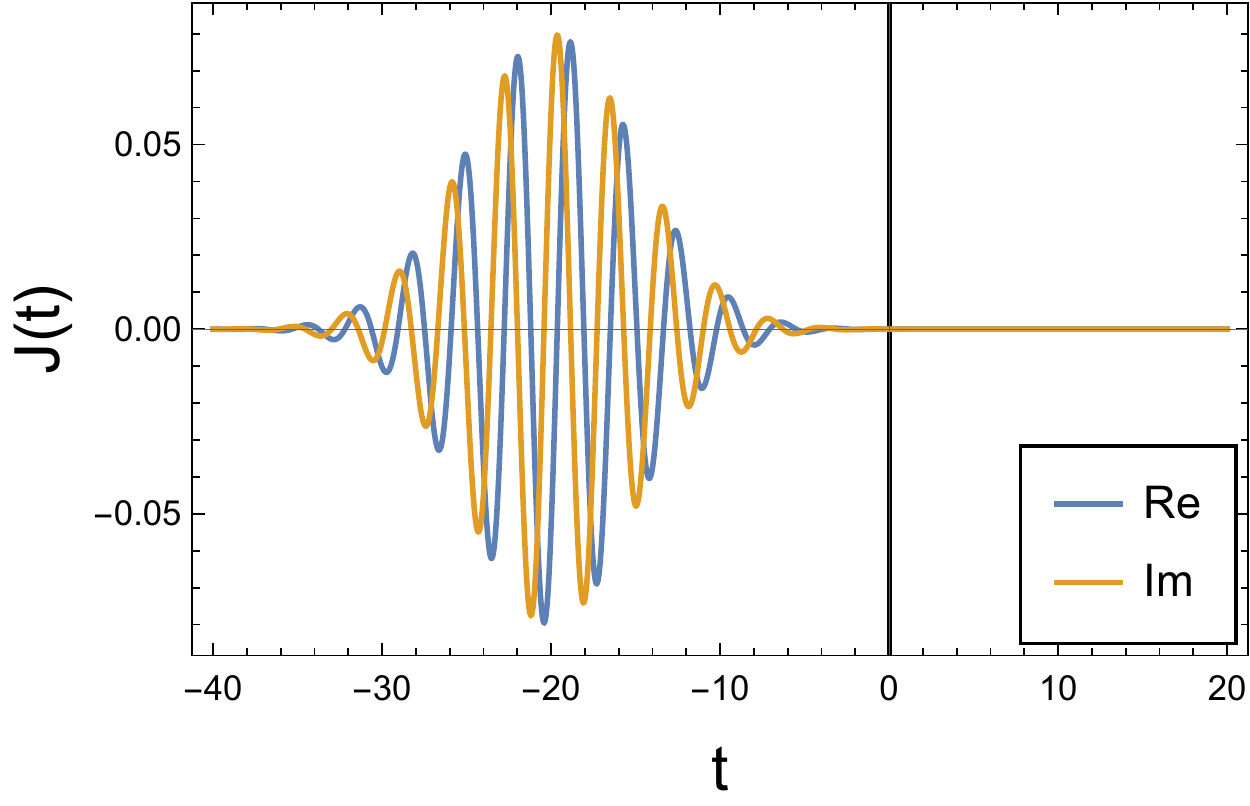}
\caption{Time dependence of the source $J(t)$,}
\label{fig:source}
\end{figure}

Substituting Eq.(\ref{Jtilde}) into Eq.(\ref{eq:sin_integral}), we obtain
\begin{equation}
\Psi(t,x) = \int_{-\infty}^{\infty} dt' J(t') \int_{-\infty}^{\infty} \frac{d\omega}{2\pi} \, \frac{\sin \omega x}{\sin \pi \omega} e^{-i \omega (t-t')} \ .
\label{Psitoy}
\end{equation}
We used the closed contour for 
the $\omega$-integration as in Fig.\ref{fig:contour0}. 
For $t-t'>0$ and $t-t'<0$, we employ the lower and upper semicircles, respectively, to make the contribution from the circular contours equal to zero.
The contribution from the infinite poles leads to
\begin{multline}
\Psi(t,x) = -\frac{i}{\pi}\int_{-\infty}^t dt' \\
\times J(t') \sum_{n'=-\infty}^\infty (-1)^{n'} \sin n' x \, e^{-in'(t-t')}\ .
\end{multline}
In this expression, the initial condition~(\ref{waveini}) is apparently satisfied.
If we use an integral contour passing the lower sides of some poles, we obtain the contribution of the poles, even at $t=-\infty$.
For a sufficiently late time $t\to\infty$,
the solution is given as 
\begin{multline}
\Psi(t,x) \simeq -\frac{i}{\pi} \sum_{n'=-\infty}^\infty (-1)^{n'} \tilde{J}(n') e^{-in't} \sin n' x \ ,
\end{multline}
where we used Eq.(\ref{Jtilde}).
If $\tilde{J}(\omega)$ is a localized function at $\omega=n$, we can realize the single $n$-th normal mode at a later time.
Such a localized function in the frequency domain can be realized by the Gaussian:
\begin{align}
&J(t) = \frac{1}{\sqrt{2\pi \sigma_t^2}} \exp\left[ -\frac{(t-T)^2}{2\sigma_t^2} - i n t \right]\ ,\\
&\tilde{J}(\omega) = \exp\left[ -\frac{\sigma_t^2}{2}(\omega-n)^2 + i(\omega-n) T \right].
\end{align}
where $\sigma_t$ and $T$ are the width and centre of the Gaussian, respectively, in the time domain.
For $\sigma_t\gg 1$, we obtain the single $n$-th normal mode at a later time. 
We take the parameters $T<0$ and $1 \ll \sigma_t \ll |T|$ such that the source is already negligible at $t=0$.
Fig.~\ref{fig:source} shows the source function for $T=-20, \sigma_t=5$, and $n=2$. 

\section{How to excite a single (quasi) normal mode in the asymptotically AdS spacetime}
\label{Howexcite}

We apply the above argument to the scalar field in asymptotically AdS spacetime. 
The general solution of Eq.(\ref{eq:schrodinger}) is given by
\begin{equation}
\Psi_{l'm'}(t,x)=\int_{-\infty}^{\infty} \frac{d\omega}{2\pi} C_{l'm'}(\omega) F_{l'}(\omega,x) e^{-i \omega t} \ ,
\end{equation}
where $F_{l'}(\omega,x)$ is the solution of 
\begin{align}
\left(\partial_x^2-V_{l'}(x)+\omega^2\right)F_{l'}(\omega,x)=0\ .
\label{eq:FT_schrodinger}
\end{align}
Note that $F_{l'}(\omega,x)$ does not depend on $m'$ because the 
above equation does not contain $m'$. 
At the horizon or the centre of the global AdS, we impose
\begin{equation}
F_{l'}(\omega,x) \sim 
\begin{cases}
e^{-i\omega x} & (r_h>0)\\
r^{l'} & (r_h=0)
\end{cases}\ ,\quad (r\to r_h)\ .
\end{equation}
Because of the linearity of the equation, there is ambiguity in choosing the overall factor of $F_{l'}(\omega,x)$. 
Here, we require
\begin{equation}
F_{l'}(\omega,x) \simeq r^{-\Delta_- +1}\ ,\quad (r\to \infty)\ ,
\label{Psiscale}
\end{equation}
i.e, the coefficient of the divergent part at the AdS boundary is tuned to 1.
Note that at (quasi) normal frequencies, $\omega=\omega_{nl'}$, where $\omega_{nl'}$ denotes the $n$-th (quasi) normal mode frequency with angular momentum $l'$. 
$F_{l'}(\omega,x)$ is ill-defined because the (quasi) normal modes decay as $\sim r^{-\Delta_+ +1}$ at infinity. 
For pure AdS $r_h=0$, $\omega_{nl'}$ is located on the real axis of the complex $\omega$ plane.
For $r_h>0$, however, it is in the lower half-plane. 
Because the potential barrier near the horizon becomes sufficiently high for large $\mu$ and $l'$, the decay rate (or tunnelling rate toward the horizon) of the quasi-normal mode, 
$\mathrm{Im}\left[\omega_{nl'}\right]$ is typically small in our computations. (For example, we have $|\mathrm{Im}\left[\omega_{nl'}\right]|<10^{-16}$ for the quasi-normal mode shown in Fig.\ref{fig:VandQNM}.)

For the pure AdS spacetime $r_h=0$, we obtain the exact solution as
\begin{multline} 
F_{l'}(\omega,x) = 
\frac{\Gamma(\gamma-\alpha)\Gamma(\gamma-\beta)}{\Gamma(\gamma)\Gamma(\gamma-\alpha-\beta)}\\
\times \cos^{-\nu+1/2}x\, \sin^{l+1}x\, {}_2F_1\left[ \alpha,\beta,\gamma, \sin^2x \right] 
\label{AdSexact}
\end{multline}
where $x=\arctan (r)$ and 
\begin{equation}
\begin{split}
&\alpha=\frac{1}{2}\left(\omega+l-\nu+\frac{3}{2}\right)\ ,\\
&\beta=\frac{1}{2}\left(-\omega+l-\nu+\frac{3}{2}\right)\ , 
\quad \gamma=l+\frac{3}{2}\ .
\end{split}
\end{equation}
$\Gamma$ and ${}_2F_1$ are the gamma function and the Gaussian hypergeometric function, respectively. 
Then, $F_{l'}(\omega,x)$ has poles at the normal frequencies $\omega=\omega_{nl'} = 2n+l'+\nu+3/2$. 
For $r_h>0$, numerical integration is required to determine $F_{l'}(\omega,x)$.

Applying the Fourier transformation to the boundary condition $r^{-1} \Psi(t,x)|_{r\to\infty}=J_{l'm'}(t)r^{-\Delta_-}$, we obtain $C_{l'm'}(\omega)=\tilde{J}_{l'm'}(\omega)$, and the general solution is written as
\begin{equation}
\Psi_{l'm'}(t,x)=\int_{-\infty}^{\infty} \frac{d\omega}{2\pi} \tilde{J}_{l'm'}(\omega) F_{l'}(\omega,x) e^{-i \omega t} \ ,
\label{Psigen0}
\end{equation}
where 
\begin{equation}
\tilde{J}_{l'm'}(\omega) \equiv \int_{-\infty}^{\infty} dt'\, J_{l'm'}(t') e^{i \omega t'}\ .
\label{Jlmtilde}
\end{equation}
Again, note that the integral contour should pass the upper side of the poles for the initial condition $\Psi_{l'm'}(t,x)|_{t\to-\infty}=0$.
Substituting Eq.(\ref{Jlmtilde}) into Eqs.(\ref{Psigen0}) and taking the consistent closed contour for the $\omega$ integration in a manner similar to Fig.\ref{fig:contour0}, we obtain
\begin{multline}
\Psi_{l'm'}(t,x) \\
= \int_{-\infty}^t dt' J_{l'm'}(t') \sum_{n'} q_{n'l'}(x) e^{-i\omega_{n'l'}(t-t')}\ .
\label{sumqnm}
\end{multline}
where
\begin{equation}
q_{nl'}(x)\equiv -i (\omega-\omega_{nl'})F_{l'}(\omega,x) |_{\omega\to \omega_{nl'}}\ ,
\label{qdef}
\end{equation}
is the quasi-normal mode function.
For a sufficiently late time, this solution becomes
\begin{equation}
\Psi_{l'm'}(t,x) = \sum_{n'} \tilde{J}_{l'm'}(\omega_{n'l'}) q_{n'l'}(x) e^{-i\omega_{n'l'}t}\ .
\label{polesum}
\end{equation}
Thus, if $\tilde{J}_{l'm'}(\omega)$ is localized at the (quasi) normal frequency, we have a single (quasi) normal mode later.
This situation is realized by
\begin{align}
&J_{l'm'}(t) = \frac{A_{l'm'}}{\sqrt{2\pi\sigma_t^2}} \exp\left[ -\frac{(t-T)^2}{2\sigma_t^2}-i \omega_{nl'} t \right],\label{eq:source}\\
&\tilde{J}_{l'm'}(\omega) = A_{l'm'} \exp\bigg[ -\frac{\sigma_t^2}{2}(\omega-\omega_{nl'})^2 \nonumber\\
&\hspace{4cm}+ i(\omega-\omega_{nl'}) T \bigg]\label{eq:source2}\ ,
\end{align}
where $A_{l'm'}$ is a complex constant.

Let us denote $x=x_0$, at which $|q_{nl}(x)|$ takes the maximum value.
We choose the overall constant of the source as
\footnote{
For pure AdS spacetime $r_h=0$, we choose the overall constant of the source as
\[
    A_{l'm'}
    =\Gamma(\nu)\sqrt{\frac{\left(2n+l'+\nu+3/2\right)\Gamma(n+1)\Gamma(n+l'+3/2)}{2\Gamma(n+1+\nu)\Gamma(n+l'+3/2+\nu)}}
\]
instead of Eq.\eqref{Adef} in the actual calculation.
This leads to an ordinary normalization condition $\langle Q_{nl'}|Q_{n'l'}\rangle=\delta_{nn'}$
}
\begin{equation}
A_{l'm'}=\frac{1}{q_{n'l'}(x_0)}\ .
\label{Adef}
\end{equation}
Then, for a sufficiently large $\sigma_t$, Eq.(\ref{polesum}) becomes
\begin{equation}
\Psi_{l'm'}(t,x) \simeq Q_{n'l'}(x) e^{-i\omega_{n'l'}t}\ ,
\label{PsiQ}
\end{equation}
where 
\begin{equation}
Q_{n'l'}(x)=\frac{q_{n'l'}(x)}{q_{n'l'}(x_0)}\ .
\label{Qdef}
\end{equation}
This is the quasi-normal mode function, which is normalized such that its peak is equal to 1, as shown in Fig.~\ref{fig:VandQNM}.

\section{Numerical method for time evolution}

\begin{figure*}[htbp]
\includegraphics[width=150mm]{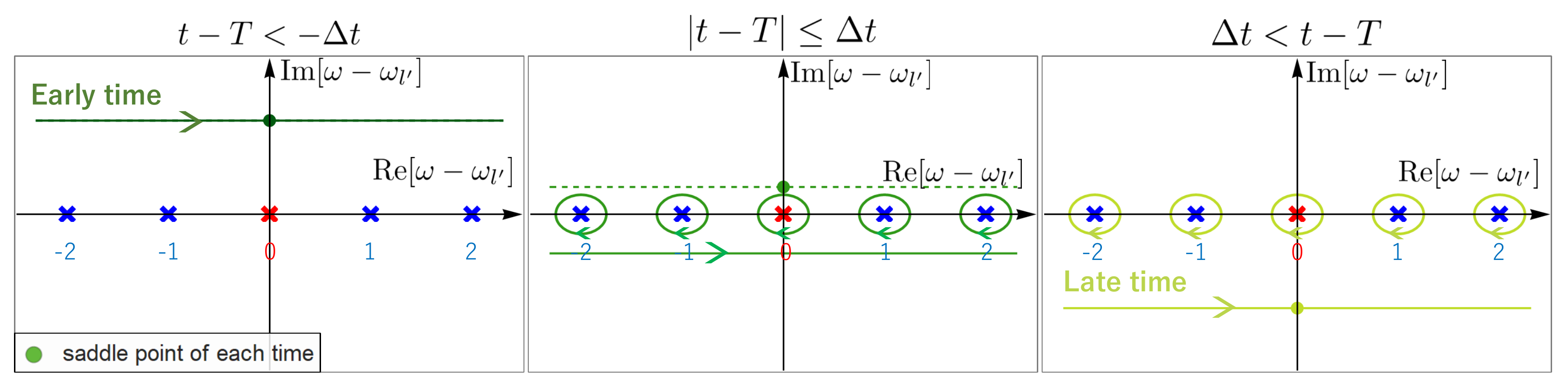}
\caption{\label{fig:contour}Time dependence of integral contour}
\end{figure*}

In section \ref{Howexcite}, we found that the single (quasi) normal mode is created by the source~(\ref{eq:source}) and (\ref{eq:source2}).
Numerical calculations are necessary to determine the time evolution of the scalar field at intermediate times.
In our numerical calculation, we evaluate the $\omega$-integral of Eq.(\ref{Psigen0}). 
(Although Eq.(\ref{sumqnm}) is an equivalent expression, we find that the convergence of summation $n'$ is slow, and Eq.(\ref{Psigen0}) is better for numerical calculations.)
Hereafter, we focus only on the creation of the fundamental tone $n=0$. For notational simplicity, we omit the index ``$0$'' for the fundamental tone, as
\begin{equation}
\omega_{0l'}=\omega_{l'}\ ,\quad q_{0l'}(x)=q_{l'}(x)\ ,\quad Q_{0l'}(x)=Q_{l'}(x)\ .
\end{equation}
Substituting the explicit expression of the source~(\ref{eq:source2}) into Eq.(\ref{Psigen0}) and completing the square with respect to $\omega$ in the exponent, we obtain:
\begin{multline}
\Psi_{l'm'}(t,x)=\\
\sqrt{2\pi\sigma_t^2} J_{l'm'}(t) \int_{-\infty}^{\infty} \frac{d\omega}{2\pi}
e^{-\frac{\sigma_t^2}{2}\left(\omega-\omega_{l'}+i\tau\right)^2} F_{l'}(\omega,x) \ ,
\label{Psiint}
\end{multline}
where 
\begin{equation}
\tau\equiv \frac{t-T}{\sigma_t^2}\ .
\end{equation}
After completing the square, the remaining terms are gathered outside the integral and can be written simply by the source in the time domain~(\ref{eq:source}).
The contour in this integral is on the real axis of the complex $\omega$plane. 
However, this contour is unsuitable for actual numerical integration. The integrand is proportional to
\begin{equation}
e^{-\frac{\sigma_t^2}{2}z^2} e^{-i(t-T)(\omega-\omega_{l'})} F_{l'}(\omega,x)\ . 
\end{equation}
When $|t-T|$ is large, this integrand quickly oscillates as a function of $\omega$, and the numerical integration loses its accuracy.
In addition, $F_{l'}(\omega,x)$ changes rapidly when $\omega$ is close to the (quasi) normal frequency. (Note that the imaginary part of the quasi-normal frequency is typically very small, even for $r_h>0$.)
It is possible to suppress the oscillation of the integrand by changing the contour 
to pass through the saddle point of the exponential factor $\omega=s+\omega_{l'}-i\tau$ ($-\infty < s <\infty$).
However, when $|\tau|$ is small, this contour passes near the poles of $F_{l'}(\omega,x)$.
Hence, we slightly modify the contour for a small $|\tau|$ as 
\begin{align}
\omega-\omega_{l'} = 
\begin{dcases}
s-i \tau &\mathrm{for}\ \ |t-T|>\Delta t\\
s-i\Delta t/\sigma_t^2 &\mathrm{for}\ \ |t-T|\leq \Delta t
\end{dcases}\ ,
\label{Ct}
\end{align}
where $s$ is the parameter of the contour integration and 
$\Delta t$ is an artificial parameter for the period to avoid poles. (In our numerical calculation, we used $\Delta t= \sigma_t$.)
In Fig.\ref{fig:contour}, we show the time-dependent integral contour in the complex plane of $\omega-\omega_{l'}$. 
The red and blue crosses are the poles of the fundamental tone and the overtones, respectively, which exist on or below the real axis. When $|t-T|>\Delta t$, the contour passes the saddle point depicted by the green dot, as shown in the left and right figures, respectively. In the middle figure, we show the contours for $|t-T|\leq \Delta t$. The contour is given by a fixed path to avoid poles and does not pass through the saddle point, which is depicted by a dashed line.
We should add the contributions of the poles when $-\Delta t \leq t-T$, because the contour goes under the poles. From Eq.(\ref{Psigen0}), the contribution of the poles is:
\begin{equation}
\begin{split}
\Psi^\textrm{pole}_{l'm'}(t,x)&= \sum_{n'}\tilde{J}_{l'm'}(\omega_{n'l'}) q_{n'l'}(x) e^{-i \omega_{n'l'} t} \\
&\simeq \tilde{J}_{l'm'}(\omega_{l'}) q_{l'}(x) e^{-i \omega_{l'} t}\\
&\simeq Q_{l'}(x) e^{-i \omega_{l'} t}\ ,
\end{split}
\label{Psigen}
\end{equation}
where $q_{l'}(x)$ and $Q_{l'}(x)$ are the (quasi) normal-mode functions defined in Eqs.(\ref{qdef}) and (\ref{Qdef}).
At the second equality, we neglect the poles of the overtone modes because they are suppressed by $\tilde{J}_{l'm'}(\omega)$. 
For the final equality, we used Eqs.(\ref{eq:source2}) and (\ref{Adef}), respectively.
In summary, the expression of the scalar field suitable for numerical evaluation is given by
\begin{multline}
\Psi_{l'm'}(t,x)=\\
\sqrt{2\pi\sigma_t^2} J_{l'm'}(t) \int_{C_t} \frac{d\omega}{2\pi}
e^{-\frac{\sigma_t^2}{2}\left(\omega-\omega_{l'}+i\tau\right)^2} F_{l'}(\omega,x)\\
+ Q_{l'}(x) e^{-i \omega_{l'} t}\,\theta(t-T+\Delta t)\ ,
\label{Psiint2}
\end{multline}
where $C_t$ denotes the contour in Eq.(\ref{Ct}) and $\theta$ is a step function.

We consider discrete points on the contour parameters as $s=s_k\equiv -5/\sigma_t + k \Delta s$, where $\Delta s = 0.1/\sigma_t$ and $k=0,1,\cdots, 100$.
For each $s_k$, the complex value of $\omega=\omega_k$ is determined using Eq.(\ref{Ct}). 
For $\omega=\omega_k$, we integrate Eq.(\ref{eq:FT_schrodinger}) from the horizon ($x=x_\textrm{min}\simeq -3$) to infinity ($x=x_\textrm{max}\simeq -0.01$), using the 4th-order Runge-Kutta method. 
(For $r_h=0$, we have the exact solution~(\ref{AdSexact}).)
We obtain a trial solution $F^\textrm{trial}_{l'}(\omega,x)$ by setting the boundary conditions to $F^\textrm{trial}_{l'}(\omega,x_\textrm{min})=1$ and $\partial_x F^\textrm{trial}_{l'}(\omega,x_\textrm{min})=-i\omega$. From the asymptotic behaviour of the trial solution, $F^\textrm{trial}_{l'}(\omega,x_\textrm{min})\simeq j_{l'}(\omega) r^{-\Delta_- +1}$, 
we obtain the coefficient of the divergent term $j_l(\omega)$. Subsequently, the solution that satisfies Eq.(\ref{Psiscale}) is given by $F_{l'}(\omega,x)=F^\textrm{trial}_{l'}(\omega,x)/j_{l'}(\omega)$.
We perform the integral in Eq.(\ref{Psiint2}) using the trapezoidal rule. 

To obtain the quasi-normal mode function, we tune $\omega$ using the shooting method such that the solution decays at infinity, i.e, $j(\omega)=0$. We then obtain the quasi-normal frequency as 
$\omega=\omega_{l'}$ and mode function $F^\textrm{trial}_{l'}(\omega_{l'},x)$. The overall factor of $F^\textrm{trial}_{l'}(\omega_{l'},x)$ differs from that of $q_{l'}(x)$ defined in Eq.(\ref{qdef}). Their relationship is given by:
\begin{equation}
\begin{split}
q_{nl'}(x)&=-i\, \frac{\omega-\omega_{l'}}{j(\omega)} F^\textrm{trial}_{l'}(\omega,x) \bigg|_{\omega\to \omega_{l'}}\\
&=-\frac{i}{j_{l'}'(\omega_{l'})} F^\textrm{trial}_{l'}(\omega_{l'},x) \ ,
\end{split}
\end{equation}
where $j_{l'}'(\omega)=d\, j_{l'}(\omega)/d\omega$.
We evaluate $j_{l'}'(\omega)$ using numerical differentiation, $j_{l'}'(\omega_{l'})\simeq (j_{l'}(\omega_{l'}+h)-j_{l'}(\omega_{l'}-h))/(2h)$ ($h\simeq 0.01$).

\begin{figure}[htbp]
\includegraphics[width=65mm]{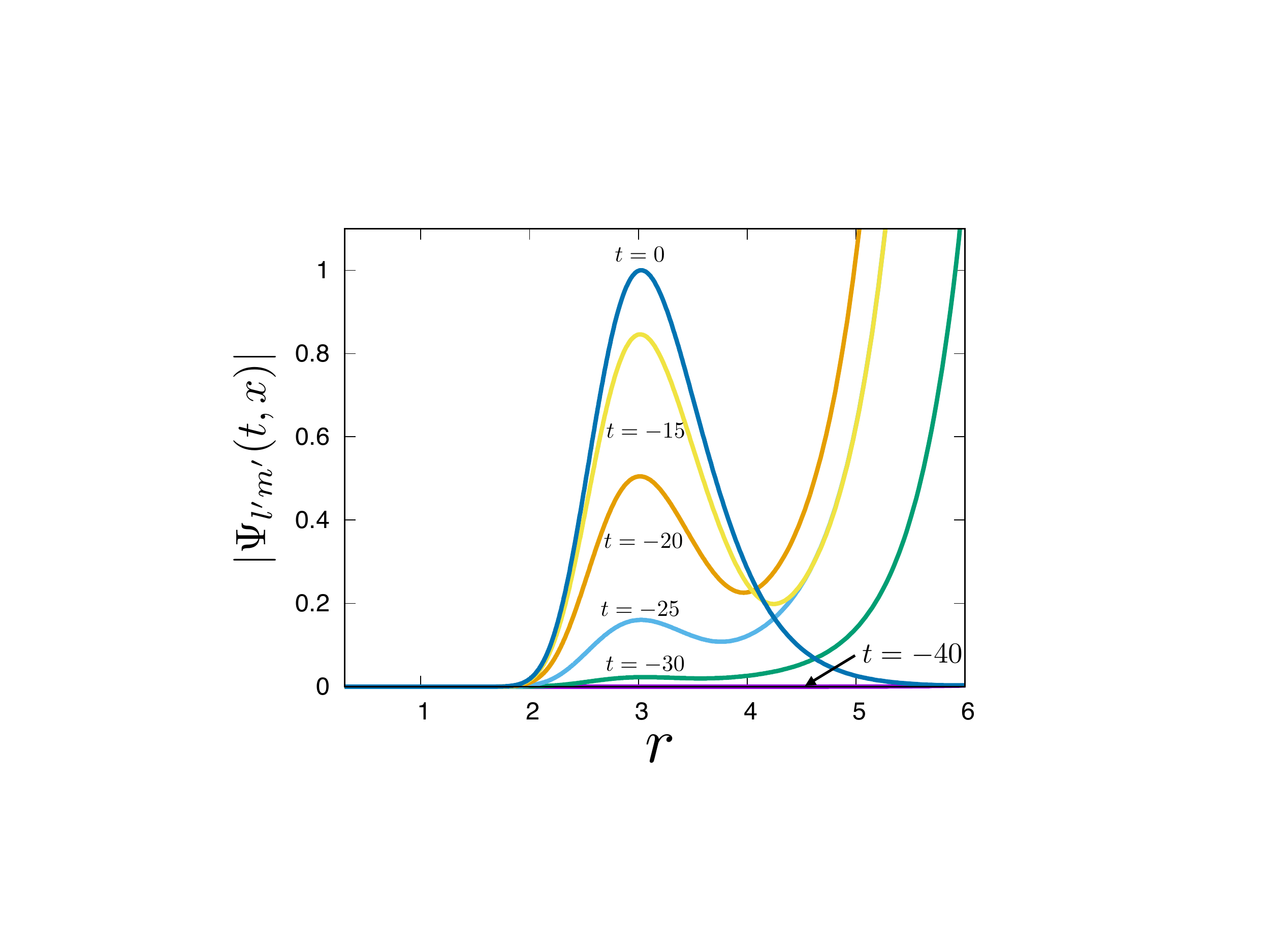}
\caption{Creation of the single quasi-normal mode using the source~(\ref{eq:source}). The parameters are $r_h=0.3, \nu=20.5$, and $l'=210$.}
\label{fig:Psi_evol}
\end{figure}

Fig.\ref{fig:Psi_evol} shows the time evolution of $\Psi_{l'm'}(t,x)$ for $r_h=0.3, \nu=20.5$ and $l'=210$.
The quasi-normal mode is created at a late time.

\section{Orbiting scalar field solution}

In the previous sections, we explained the method for constructing a single (quasi) normal mode using the boundary condition of the scalar field 
for fixed $l'$ and $m'$.
The scalar field in the coordinate space is written using their superposition:
\begin{align}
&\Psi(t,r,\theta,\phi)\equiv r\Phi(t,r,\theta,\phi) \nonumber \\
= & \sum_{l'm'}c_{l'm'}\Psi_{l'm'}(t,x)Y_{l'm'}(\theta,\phi)\ ,
\label{eq:SHdecomposition}
\end{align}
where $c_{l'm'}$ are arbitrary constants. 
Then, for a sufficiently long time, source~(\ref{eq:source}) is sufficiently small, and $\Psi_{l'm'}(t,x)$ is approximated by Eq.(\ref{PsiQ}).
Therefore, we have 
\begin{align}
\label{eq:SHdecomposition_late_1}
\Psi(t,r,\theta,\phi)\simeq \sum_{l'm'}c_{l'm'}e^{-i\omega_{l'}t} Q_{l'}(x) Y_{l'm'}(\theta,\phi)\ .
\end{align}
This expression is not functionally complete, as only fundamental tones are included.
Let us choose constants $c_{l'm'}$ so that the scalar field is localized  on the 3-dimensional time-slice at a later time and orbits in the $\phi$ direction with an angular velocity $\Omega$.
Such a configuration of the scalar field is given by 
\begin{multline}
    \Psi(t,r,\theta,\phi)
    =e^{-i\omega_mt}Q_{m}(x)\\
    \times\exp\left[-\frac{(\theta-\pi/2)^2}{2\sigma_\theta^2}-\frac{(\phi-\Omega t)^2}{2\sigma_\phi^2}+im\phi\right] ,\label{eq:answer_late_1}
\end{multline}
which is localized at $\theta=\pi/2$ and $\phi=\Omega t$, with widths $\sigma_\theta$ and $\sigma_\phi$. 
(We take the domain of the coordinate $\phi$ as $-\pi<\phi-\Omega t \leq \pi$.)
We multiply $e^{im\phi}$ by the Gaussian, which gives momentum $m$ to the scalar field: 
$\hat{L}_z \Psi(t=0,r,\theta,\phi) \simeq m \Psi(t=0,r,\theta,\phi)$, where $\hat{L}_z \equiv -i \partial_\phi$ denotes the angular momentum operator.
For parameters $\sigma_\theta,\sigma_\phi$ and $m$, we require 
\begin{align}
\frac{1}{m}\ll\sigma_\theta,\sigma_\phi\ll1\ ,
\label{eq:classical_condition}
\end{align}
to localize the scalar field in both position and momentum space. 
The angular velocity $\Omega$ will be determined later.

We now introduce $\phi_t\equiv \phi-\Omega t$. Then, \eqref{eq:SHdecomposition_late_1} can be rewritten as follows:
\begin{align}
&\sum_{l'm'}c_{l'm'} e^{-i\omega_{l'}t} Q_{l'}(x) Y_{l'm'}(\theta,\phi)\nonumber\\
&=\sum_{l'm'}c_{l'm'} e^{-i\left(\omega_{l'}-m'\Omega\right)t} Q_{l'}(x) Y_{l'm'}(\theta,\phi_t) ,\label{eq:SHdecomposition_late_2}
\end{align}
and Eq.(\ref{eq:answer_late_1}) is rewritten as
\begin{align}
&\Psi(t,r,\theta,\phi)=e^{-i\left(\omega_m-m\Omega\right)t}Q_{m}(x)g(\theta,\phi_t), \label{eq:answer_late_2} \\
&g(\theta,\phi)\equiv \exp\left[-\frac{(\theta-\pi/2)^2}{2\sigma_\theta^2}-\frac{\phi^2}{2\sigma_\phi^2}+im\phi\right] .\label{gdef}
\end{align}
From Eqs.\eqref{eq:SHdecomposition_late_2} and \eqref{eq:answer_late_2}, the equation to determine $c_{l'm'}$ is
\begin{multline}
c_{l'm'} Q_{l'}(x) e^{-i\left(\omega_{l'}-m'\Omega\right)t} \\
\simeq \langle Y_{l'm'}|g \rangle Q_{m}(x) e^{-i\left(\omega_{m}-m\Omega\right)t}\ ,
\label{ccond}
\end{multline}
where $\langle f_1 | f_2 \rangle \equiv \int d\Omega f_1^\ast(\theta,\phi) f_2(\theta,\phi)$.
As we will see shortly, this equation is approximately satisfied, even though Eq.(\ref{eq:SHdecomposition_late_2}) is not complete.
Because of the condition $\sigma_\theta,\sigma_\phi \ll 1$, 
the inner product $\langle Y_{l'm'}|g \rangle$ is explicitly calculated as follows.
When $l'-m'$ is an even number, we obtain~\footnote{
When $l'-m'$ is even, for $|\theta-\pi/2|\ll 1$, the associated Legendre polynomial in the spherical harmonics is approximated as 
\[
 P_l^m(\cos\theta)\simeq (-1)^{(l-m)/2}\frac{(l+m-1)!!}{(l-m)!!}\cos k (\theta-\pi/2)\ .
\]
Then, we can perform the Gaussian integral. 
When $l'-m'$ is odd, $P_l^m(\cos\theta)$ is odd function with respect to $\theta-\pi/2$ and the integral is zero.
}
\begin{align}
&\langle Y_{l'm'}|g \rangle\simeq \sigma_\theta\sigma_\phi (-)^{\frac{l'+m'}{2}} \sqrt{\pi(2l'+1)\frac{(l'-|m'|)!}{(l'+|m'|)!}}\nonumber\\
&\times \frac{(l'+|m'|-1)!!}{(l'-|m'|)!!} 
\exp\bigg[-\frac{\sigma_\phi^2}{2}(m'-m)^2\nonumber\\
&\hspace{2cm}-\frac{\sigma_\theta^2}{2}\left(l'(l'+1)-m'^2+\frac{1}{2}\right)\bigg]\ .
\label{eq:inner_Yg}
\end{align}
When $l'-m'$ is an odd number, we have $\langle Y_{l'm'}|g \rangle=0$.
This is non-negligible only when $l'\sim m'\sim m$. The widths are given by
\begin{align}
    &\Delta m'\equiv \left|m'-m\right|\simeq\frac{1}{\sigma_\phi}\ll m\ ,\label{mpm}\\
    &\Delta l'\equiv \left|l'-m'\right|\simeq\frac{1}{\sigma_\theta^2 m}\ll m\ ,\label{lpmp}
\end{align}
where inequalities follow from Eq.\eqref{eq:classical_condition}. 
We will further require 
\begin{equation}
 \frac{1}{\sigma_\theta^2 m} \lesssim 1\ .
\label{addcond}
\end{equation}
We will see that this condition is necessary so that the scalar field is localized for sufficiently long time.
(Parameters used in the main text satisfy this condition.)
Under this condition, we have 
\begin{equation}
 |l'-m|\simeq\frac{1}{\sigma_\phi}\ .
\label{lpm}
\end{equation}

Our choice of $c_{l'm'}$ is 
\begin{equation}
c_{l'm'}=\langle Y_{l'm'}|g \rangle\ .
\label{eq:clm}
\end{equation}
In this choice, we can approximate $Q_{l'}(x)\simeq Q_{m}(x)$ in Eq.(\ref{ccond}) since $c_{l'm'}$ is localized at $l'\sim m$.
We can explicitly verify that $Q_{l'}(x)$ does not radically depend on $l'$ in Fig. \ref{fig:VandQNM}.
Eq.(\ref{ccond}) also requires that the phase factors be approximately equal for a long time $T'$, i.e, 
$e^{-i\left(\omega_{l'}-m'\Omega\right)T'}\simeq e^{-i\left(\omega_{m}-m\Omega\right)T'}$. 
Taking the time as the period of the orbital motion, $T'=1/\Omega$, we can write this condition as
\begin{multline}
 \frac{1}{\Omega}|(\omega_{l'}-\omega_m)-(m'-m)\Omega|\\
\simeq \left|(l'-m)\frac{1}{\Omega}\left(\frac{\partial \omega_{l'}}{\partial l'}\right)_{l'=m}-(m'-m)  \right| \lesssim 1\ .
\label{loclizecond}
\end{multline}
Typically, $|m'-m|$ and $|l'-m|$ are much larger than $1$ as in Eqs.(\ref{mpm}) and (\ref{lpm}).
In order for above inequality to be satisfied, 
the two terms must be cancelled out with sufficiently good accuracy. 
The cancellation occurs only when 
\begin{equation}
 \Omega=\frac{\partial \omega_{l'}}{\partial l'}\bigg|_{l'=m}\ .
\end{equation}
Then, the inequality is written as $|l'-m'|\lesssim 1$ and this is satisfied because of Eqs.(\ref{lpmp}) and (\ref{addcond}).
Therefore, the choice of $c_{l'm'}$ in Eq.\eqref{eq:clm} creates the localized orbiting scalar field solution in Eq.\eqref{eq:answer_late_1}.

The source in the coordinate space is written as
\begin{equation}
\mathcal{J}(t,\theta,\phi)
=\sum_{l'm'} J_{l'm'}(t) Y_{l'm'}(\theta,\phi) \langle Y_{l'm'}|g \rangle \ ,
\end{equation}
where $J_{l'm'}(t)$ is defined in Eq.(\ref{eq:source}). From a similar argument, 
we obtain
\begin{multline}
\mathcal{J}(t,\theta,\phi) \propto \exp\bigg[-\frac{(t-T)^2}{2\sigma_t^2}\\
-\frac{(\theta-\pi/2)^2}{2\sigma_\theta^2}
-\frac{(\phi-\Omega t)^2}{2\sigma_\phi^2}
-i\omega_m t+im\phi \bigg]\ .
\end{multline}
This function is localized in $S^2$, and 
its centre rotates on the equator with angular velocity $\Omega$. 
This has a wavenumber $m$ along the $\phi$-direction and oscillates over time with frequency $\omega$.


\bibliography{MassiveStar}
